\documentclass[%
reprint,
superscriptaddress,
nofootinbib,
amsmath,amssymb,
aps,
prl,
]{revtex4-2}

\usepackage{graphicx}
\usepackage{dcolumn}
\usepackage{bm}
\usepackage{siunitx}
\usepackage{physics}
\usepackage{soul}

\begin{document}


\title{Beam Intensity Limitations in Future Multi-Bend Achromat Light Sources}

\author{I. Agapov}%
\affiliation{%
Deutsches Elektronen-Synchrotron DESY, Notkestr. 85, 22607 Hamburg, Germany 
}%

\author{S.~A. Antipov}
\affiliation{%
Deutsches Elektronen-Synchrotron DESY, Notkestr. 85, 22607 Hamburg, Germany 
}%

\date{\today}

\begin{abstract}
   We show that emittance of fourth-generation 6 GeV machines such as PETRA IV is close to what is theoretically achievable due to beam intensity limitations from space charge and intra-beam scattering. Investigating these limitations, in particular their scaling with the bare lattice emittance and the beam energy, we argue that achieving further significant emittance reduction and increase in radiation brightness is only possible by increasing the beam energy. We outline the design and technological challenges on the way to such improvement.
\end{abstract}

\maketitle


\section{Introduction}

The quest for higher photon beam brightness in light sources to advance a wide spectrum of experimental techniques ushered in the so-called fourth generation of low-emittance ring design based on multi-bend achromat (MBA) lattices \cite{Einfeld:xe5006}. Many of these machines have now become operational or will be constructed in the near future \cite{Chapman:me6227,Raimondi2023, schroer2018petra, Schroer2019}. In this latest generation a number of collective effects are encountered, that have not manifested themselves in the third-generation light sources. For both low-energy and high-energy machines the intra-beam scattering (IBS) has a significant impact on equilibrium emittance, requiring mitigation by higher-harmonic cavities \cite{MIGLIORATI1995215}. Additionally, the space charge (SC) tune shift cannot be neglected anymore. For fourth-generation light sources it exceeds the synchrotron tune and leads to a variety of dynamic effects~\cite{PhysRevAccelBeams.28.024401}.


In this paper we investigate these effects and their possible implications on further improvement of brightness.  As analytical treatment of practical low-emittance lattices appears impossible and a practical lattice design to reach certain target parameters involves multiple tradeoffs, we study these limitations through a mixture of analytical estimates and scaling of the hybrid six-bend achromat (H6BA) lattice~\cite{Agapov:2024wey, Raimondi:2023rby}, the latest concept in the multibend achromats, on which, for instance, the design of the PETRA~IV facility is based. The scaled H6BA lattice is sufficiently representative of what can be achieved when pushing the multi-bend achromats towards further compactness.


Our analysis shows that while the collective instabilities are not expected to be an obstacle on the way to lowering emittance, both space charge and IBS present severe limitations that prevent a significant emittance reduction beyond what is currently achieved. The strategy to mitigate those limitations is to increase the electron beam energy.


\newpage

\section{Natural emittance scaling}

First, let us recall the well-known formulae for equilibrium beam parameters in electron storage rings \cite{Sands:1969lzn}, \cite{Wolski:2014aab}. The natural rms emittance, energy spread and bunch length are  
\begin{equation}
    \begin{split}
        &  \epsilon_0 = C_q \gamma^2 \frac{I_5}{j_x I_2},  \\
        &  \sigma_{\delta}^2 = C_q \gamma^2 \frac{I_3}{j_z I_2},\\
        &  \sigma_{z} = \frac{\alpha_C c}{\omega_s} \sigma_{\delta},
    \end{split}
\end{equation}
 where $C_q \approx 3.8\times 10^{-13}~\text{m}$, $\alpha_C$ is the momentum compaction factor,  $\omega_s$ the synchrotron frequency, $c$ the speed of light and $\gamma$ the relativistic factor.  The so-called radiation integrals along the circumference are defined as
\begin{equation}
    \begin{split}
        & I_2 = \oint \frac{1}{\rho_2} ds , \\
        & I_3 = \oint \frac{1}{|\rho|^3} ds , \\
        & I_4 = \oint \frac{\eta_x}{\rho} \left(\frac{1}{\rho^2} + 2 k_1\right)ds ,\\
        & I_5 = \oint \frac{\mathcal{H}_x}{|\rho|^3} ds , 
    \end{split}
\end{equation}
where $\mathcal{H}_{x} = \gamma_x \eta_x^2 + 2 \alpha_x \eta_x \eta_{px} + \beta_x \eta_{px}^2$ is the horizontal dispersion invariant. The 
partition numbers $j_x = 1 - {I_4} / {I_2}$, $j_y=1$ and $j_z = 1 + {I_4} / {I_2}$ are influenced primarily by focusing in the combined-function magnets and indicate how the damping is shared between planes, with the damping times $\tau_{x,y,z}$ and rates $\alpha_{x,y,z}$ defined as
\begin{equation}
\tau_{x,y,z} = \alpha_{x,y,z}^{-1} = 
\frac{2}j_{x,y,z} \frac{E_0}{U_0} T_0, 
\end{equation}
where $T_0$ is the revolution period, $E_0$ the beam energy and 
\begin{equation}
    U_0 = \frac{C_{\gamma}}{2 \pi} E_0^4 I_2 
\end{equation}
is the energy loss per turn with $C_{\gamma} \approx 8.8 \times 10^{-5}$ m / $\mathrm{GeV^3}$.

We will consider a machine of fixed circumference, for example that of PETRA (2304~m), and the potential emittance reduction through low-emittance lattice beyond what will be achieved at PETRA~IV. For all multi-bend achromat lattices, including those using damping wigglers, $I_2$ is relatively independent of the machine optics. The biggest effect of emittance reduction is achieved through reducing of the $I_5$ integral by stronger focusing. The $I_3$ integral is also only weakly dependent on the choice of optics, and the equilibrium energy spread becomes a function of energy mostly. Bunch length is reduced for low-emittance lattices as it is proportional to the momentum compaction. 

\begin{figure}
    \centering
    \includegraphics[width=0.99\linewidth]{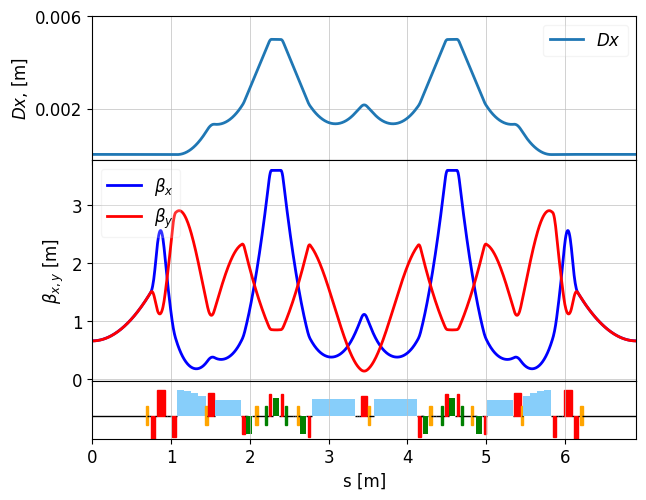}
    \caption{The scaled H6BA cell of PETRA~IV reduced in length by a factor of 4.}
    \label{fig:h6ba}
\end{figure}

Emittance reduction is usually achieved by increasing the number of bends and focusing elements in a cell of fixed length, for example, by going from the double-bend achromat (DBA) to the seven-bend achromat (7BA or H7BA) cell. Analytical expressions exist that describe emittance scaling with the number of magnets per cell for standard arrangements such as FODO or TME (theoretical minimum emittance cell). However, a practical design does not follow such cell structures, since additional constraints such as space for insertion devices should be satisfied. An example of such a design is the H6BA cell adopted for PETRA~ \cite{Agapov:2024wey, Raimondi:2023rby}. Since analytical expressions for optical functions in an MBA optics are not possible, we will study the scaling of parameters based on the scaling of the H6BA cell of PETRA~IV (Fig.~\ref{fig:h6ba}). While a realistic design of a smaller emittance lattice will certainly not follow this path, the achieved values of beta functions, dispersion invariant $\mathcal{H}$ and the radiation integrals will necessarily be similar to an H6BA cell scaled to achieve similar emittance values. The scaling is achieved by reducing all element lengths by a factor $f$, increasing quadrupole strength by $f^2$ and sextupole strength by $f^4$ to have the same corrected chromaticity. We ignore the higher order multipole (octupole) scaling as they are not essential for this analysis.



\begin{figure}[h!]
    \centering
    \includegraphics[width=1\linewidth]{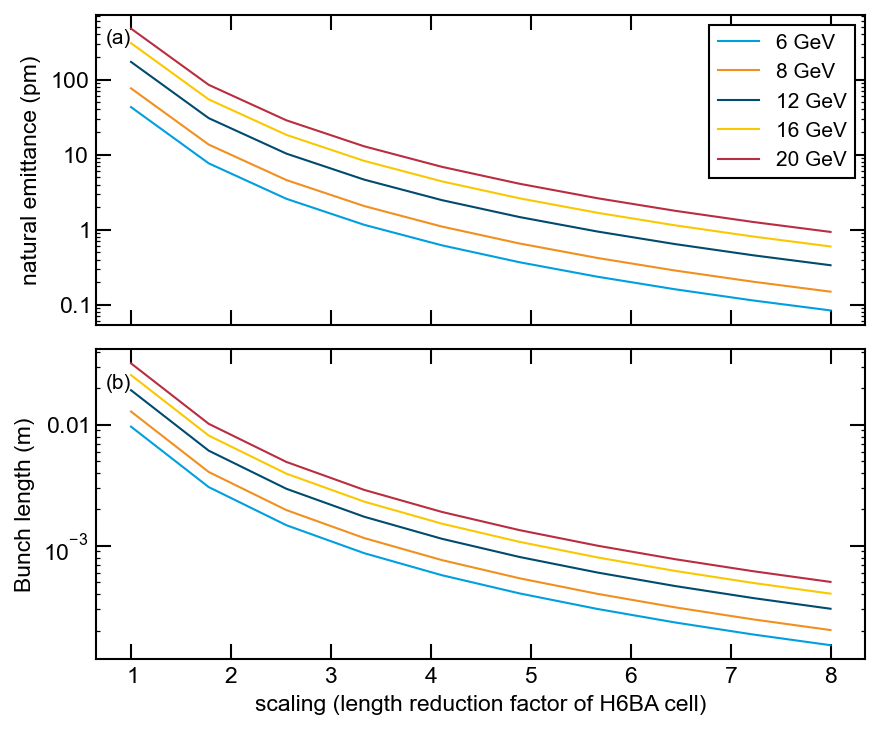}
    \caption{Scaling with energy and cell length of natural emittance (a) and bunch length (b).}
    \label{fig:emittance-scaling}
\end{figure}

\begin{figure}
    \centering
    \includegraphics[width=0.99\linewidth]{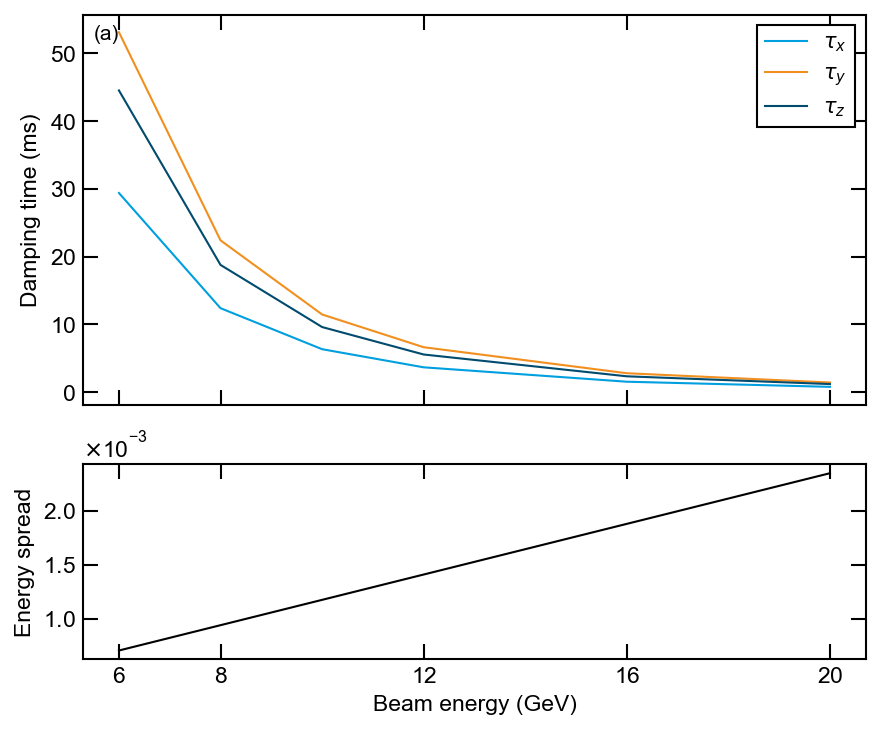}
    \caption{Damping times (a) and energy spread (b) vs. beam energy. There is no dependence on lattice scaling.}
    \label{fig:damping}
\end{figure}

Figure \ref{fig:emittance-scaling}a shows emittance as a function of scaling parameter for various beam energies.\footnote{Note that in practice emittance can be further reduced by damping wigglers (as it will be done in PETRA~IV, for example). The choice of damping wiggler parameters is usually done such that equilibrium bunch length and energy spread are minimally affected, but emittance is reduced via increase in $I_2$. $I_2$, as mentioned before, is independent on lattice scaling. 
We ignore this option in the present analysis. It does not affect the space charge limitations, but can be used to mitigate the intra-beam scattering effect to some extent, e.g. by a factor of 2 in the case of PETRA~IV, at the cost of increased radiation losses and rf power. Use of damping wigglers is usually restricted by available space and is only efficient in machines such as PETRA~IV which are built in old collider tunnels and where sections unused by insertion devices are available.} Figure \ref{fig:damping} shows the dependencies of the damping times and the equilibrium energy spread on the beam energy. They are both independent of the lattice scaling.  Figure \ref{fig:emittance-scaling}b shows the bunch length scaling with energy and lattice scaling. Here an implicit dependency of synchrotron frequency on the lattice parameters via momentum compaction $\alpha_C$ has been removed by fixing the synchrotron frequency at 1 kHz. An alternative assumption could be that the bunch length remains constant, which would however imply that the synchrotron frequency would approach zero when emittance is reduced, which is detrimental for collective instabilities. 
Both limiting cases are examined later numerically, showing minor differences in achievable emittances.

\section{Beam intensity scaling}

\subsection{Space charge estimates}

Typically, SC effects are most prominent in the frontier high-intensity hadron machines and their injectors, while electron rings benefit from the much smaller particle mass as the SC interaction decreases with the Lorentz factor. Yet, the SC effects have already been found important in the linear collider damping rings where a combination of a relatively low beam energy and a small emittance makes SC interaction significant ~\cite{Decking:2000ka, Xiao:2007lff, Rumolo:2008zzb, ZampetakisPRAB2024}. State-of-the-art light sources also share this combination, as pointed out in~\cite{PhysRevAccelBeams.28.024401}.

The Coulomb interaction between the particles will change their natural frequencies of oscillation around the reference orbit, causing the so-called space charge tune shift. In the linear model the tune shift of a Gaussian bunch is
\begin{equation}\label{eq:SC_general_case}
    \Delta \nu^{SC}_{x,y} = - \frac{N r_e C}{(2\pi)^{3/2}\gamma^3\sigma_z}
    \biggl \langle 
    \frac{\beta_{x,y}}{\sigma_{x,y}(\sigma_x+\sigma_y)} 
    \biggr \rangle,
\end{equation}
where $N$ is the number of particles per bunch, $\sigma_{x,y,z}$ are the rms horizontal, vertical and longitudinal beam sizes, $\gamma$ is the Lorentz factor, $\beta_{x,y}$ are the transverse Twiss optics functions, and $\langle ... \rangle$ denotes averaging over the ring circumference $C$. $r_e = e^2 / m_ec^2 \approx 2.8\times10^{-13}$~cm (in cgs units) stands for the classical electron radius and $c$ is the speed of light. For realistic light source lattices this formula can be further simplified and an estimate of the tune shift (here in the vertical $y$-plane, where it is usually the largest) becomes (see App.~A\ref{ap:A} for derivation)
\begin{equation}\label{eq:SC_limit}
    \Delta \nu^{SC}_{y} \approx - \frac{1}{4\kappa}\frac{N r_e C}{ (2\pi)^{3/2}\gamma^3\sigma_z\epsilon_x},
\end{equation}

where $\kappa$ is the betatron coupling ratio. As SC drives the incoherent tunes to cross the integer resonance it establishes a hard, first-principle limit on the maximum beam intensity that can be stored in the ring. Defining the peak beam brightness as $B = 2 I / \pi^2 \epsilon_x \epsilon_y =  2 I / \pi^2 \epsilon_x^2 \kappa$, where $I$ stands for the peak bunch current, its achievable upper limit in any light source then becomes
\begin{equation}\label{eq:brighness_limit}
    B \times \epsilon_x \lesssim 8 \gamma^3 I_A / \pi C,
\end{equation}
where $I_A = m_e c^3 / e \approx 17$~kA is the Alfven current.
Note that this limit is independent of the details of the focusing lattice and only depends on two factors: the beam energy and the ring circumference. For our 6~GeV 2.3-km-long PETRA~IV example with $\epsilon_x = 20$~pm one obtains that the brightness cannot exceed $B < 1.5\times10^{24}$~A/m$^2$. 
For the PETRA IV circumference, bunch length, and 6 GeV energy, the space charge effect would limit the emittance to about $5-10$ pm \cite{IPAC-2025-SC}.\footnote{The exact limit depends on bunch length and coupling ratio} The scaling of SC implies that achieving emittances of 1 pm and below should be possible by raising the energy to about 10 GeV.

\subsection{IBS estimates}

With damping rate $\alpha$ and quantum diffusion rate $D$, the beam distribution evolution is described by the Fokker-Planck equation in each of the $x,y,z$ dimensions
\begin{equation}
\partial_{t} f(J) = \alpha \partial_J f(J) + D \partial_{JJ} f(J) 
\end{equation}
If one looks for a stationary solution in the form $f_0 e^{-J/\epsilon}$ where $\epsilon$ is the equilibrium emittance, one gets $\epsilon = D / \alpha$.
Now, the damping rate is the sum of synchrotron radiation damping and IBS-induced growth, $\alpha = \alpha_{SR} - \alpha_{IBS}$, so we can write the following
\begin{equation}\label{eq:equilibrium_emittace}
\epsilon = \frac{D/\alpha_{SR}}{1 - \alpha_{IBS}/ \alpha_{SR}} = \frac{\epsilon_0}{1 - \alpha_{IBS}/ \alpha_{SR}},
\end{equation}
where $\epsilon_0$ is the equilibrium emittance without taking IBS into account \cite{Bartolini:2022bor, Bane:2002sr, Piwinski:1974it}. It is clear that for the beam to achieve equilibrium one should have $\alpha_{IBS} < \alpha_{SR}$.
For IBS growth rate not exceeding synchrotron damping, the equilibrium emittance can be found numerically by solving Eq.~(\ref{eq:equilibrium_emittace}) with emittance-dependent IBS growth rates given by
\begin{equation}
    \begin{split}
        & \alpha_{x,y}^{IBS} \approx \frac{r_e^2 c N L_C}{16 \gamma^3 \epsilon_{x,y}^\frac{7}{4} \epsilon_{y,x}^\frac{3}{4} \sigma_z \sigma_p} \left< \frac{\mathcal{H}_{x,y} \sigma_H g\left(x \right)} {\left(\beta_x\beta_y\right)^{\frac{1}{4}}}\right>,\\
        & \alpha_{p}^{IBS} \approx \frac{r_e^2 c N L_C}{16 \gamma^3 \epsilon_{x}^\frac{3}{4} \epsilon_{x}^\frac{3}{4} \sigma_z \sigma_p^3} \left< \frac{\sigma_H g\left(x \right) }{\left(\beta_x\beta_y\right)^{\frac{1}{4}}}\right>,
    \end{split}\label{eq:ibs_rates}
\end{equation}
where $L_C$ is the Coulomb log and $g(x)\approx2x^{0.021-0.044 \ln x}$. Here $\mathcal{H}_{x,y}$ are the dispersion invariants in the horizontal and the vertical plane accordingly and $\sigma_p$ is the relative rms momentum spread; the other parameters of the model are defines as 
\begin{equation}
    \begin{split}
        & x = b_x / b_y, \\
        & b_{x,y} = \sigma_H / \gamma \times \sqrt{\beta_{x,y} / \epsilon_{x,y}}, \\
        & \sigma_H = \sqrt{\sigma_p^{-2} + \mathcal{H}_x / \epsilon_x + \mathcal{H}_y / \epsilon_y }.
    \end{split}
\end{equation}

\begin{figure}
    \centering
    \includegraphics[width=1\linewidth]{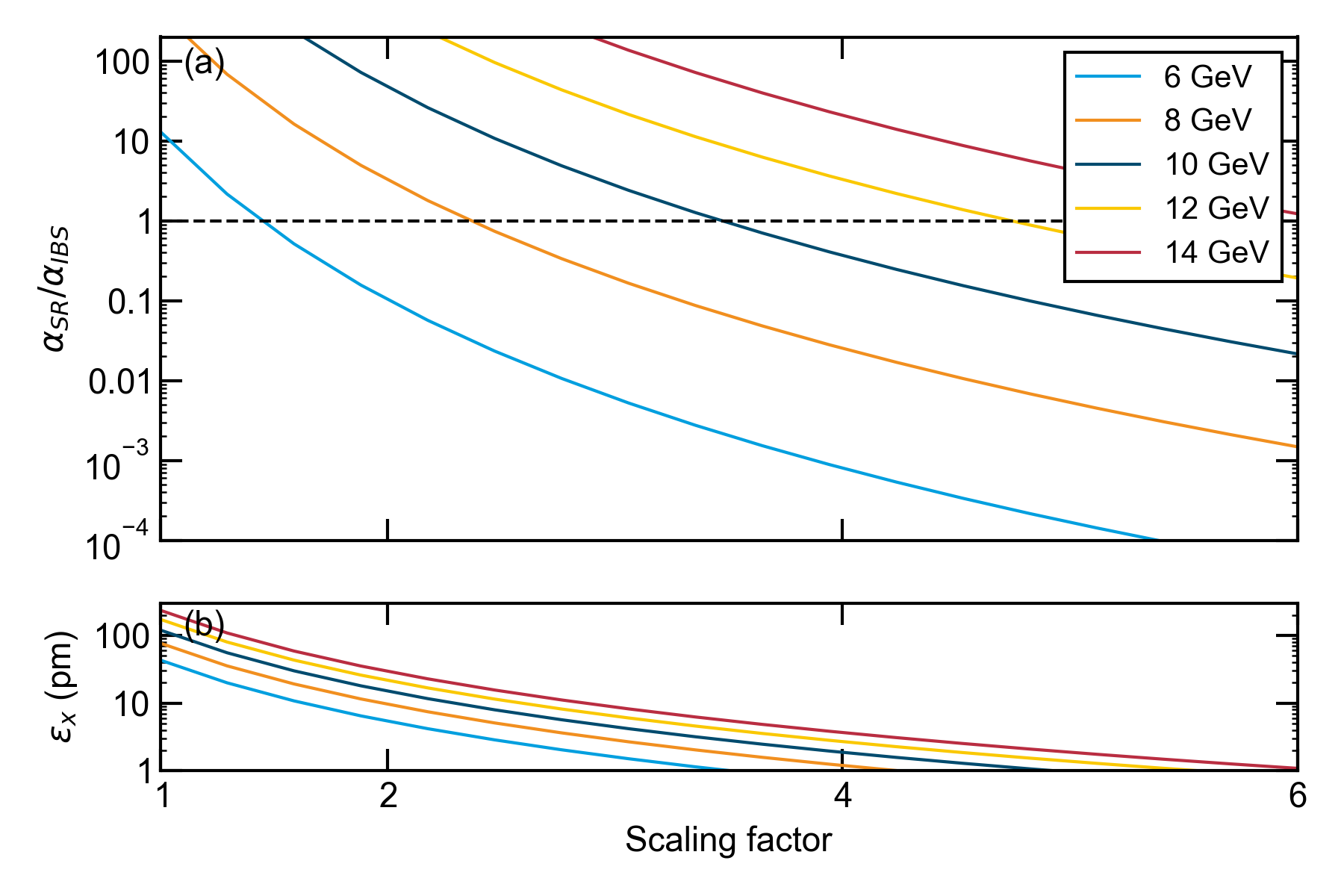}
    \caption{Evolution of SR and IBS rates (a) and emittance (b) a function of 6BA scaling. As emittance decreases with the scaling, the IBS rate exceeds the SR (dashed line) limiting the further reduction of emittance. Coupling $\kappa = 0.1$.}
        \label{fig:ibs_emittance}
\end{figure}

With the increase of the scaling factor (number of bends per cell) the equilibrium emittance and bunch length decrease (Fig.~\ref{fig:emittance-scaling}). This, in turn, enhances the IBS effect and the ratio of IBS growth rate to SR damping rate increases (Fig.~\ref{fig:ibs_emittance}) until at a certain point it becomes 1. Beyond that point the equilibrium emittance is fully dominated by the IBS. This IBS threshold is increased and smaller equilibrium emittances can be reached at higher beam energies.

\subsection{Minimum achievable emittance}

In order to find the equilibrium emittance in the presence of strong IBS we solved for an equilibrium between IBS, SR and quantum excitation given by Eq.~(\ref{eq:equilibrium_emittace}) numerically, accounting for both the longitudinal and the transverse degrees of freedom. In the latter we assumed that $\mathcal{H}_y \approx 0$ and the vertical IBS growth rate can be neglected, with the vertical emittance governed by a small betatron coupling ratio $\kappa$. The bunch length is strongly influenced by the choice of rf parameters, including the harmonic rf, and longitudinal impedance. For example, while the natural bunch length of PETRA IV is $\sim 7$~ps, these effects increase it to about 40~ps, comparable with the present PETRA III. Thus, we assumed that the natural bunch length is further increased by a factor of 10 for the calculations. Bunch charge of 1 nC for the standard brightness mode fill pattern was used.

\begin{figure}[h!]
    \centering
    \includegraphics[width=1\linewidth]{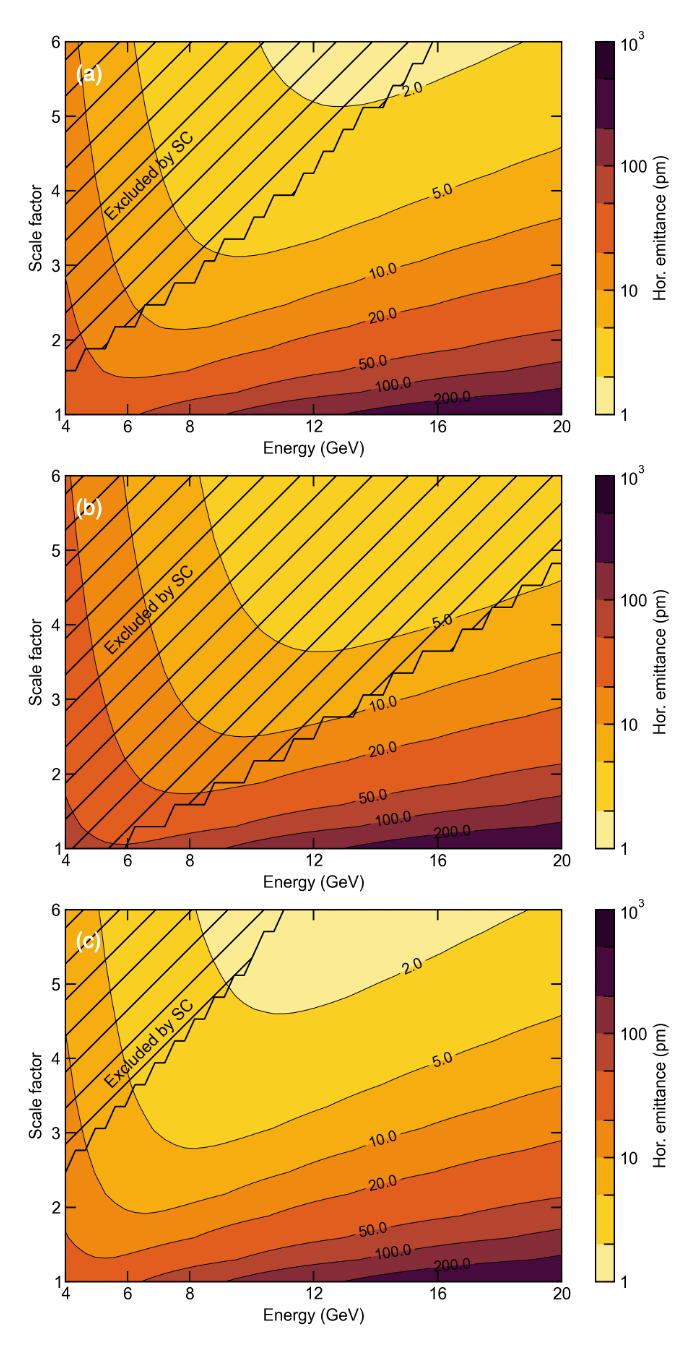}
    \caption{Achievable emittances as a function of beam energy and 6BA scaling for different couplings: $\kappa = 0.01$ (a), $\kappa = 0.1$ (b), and $\kappa = 0.5$ (c). Equilibrium emittance is defined by the balance of SR, QE, and IBS. A factor 10 bunch lengthening by a harmonic rf system is assumed. The hatched area denotes the parameters space unavailable due to the space charge tune spread. Bunch charge 1 nC.}
        \label{fig:minimum_emittance}
\end{figure}

The resulting emittance landscape for different energies and lattice scalings is presented in Fig.~\ref{fig:minimum_emittance}. Emittance increases with energy, as expected. It decreases with the scaling factor up to a point where it becomes limited by IBS, beyond which further increase of the scaling factor only slightly improves the emittance. For example, for a 6 GeV lattice the minimum emittance reaches 10~pm at a scaling factor about 2.5 (down from the nominal of $\sim 40$ pm for a H6BA) and a further reduction is possible only with a significant increase of the number of bends in the achromat.

Examining Fig.~\ref{fig:minimum_emittance} further one can notice that there seem to be optimal combinations of scale and energy for any given target emittance: the corner points on curves of equal emittance. They represent local minima of emittance with respect to variation of energy and scale. 

Further, one can see in Fig.~\ref{fig:minimum_emittance} the space charge exclusion area. As discussed above, it arises from crossing of the integer resonance by the SC tune spread and can be estimated using Eq.~(\ref{eq:brighness_limit}). The SC effectively prohibits reaching very small emittances and dictates that a decrease of the beam emittance must be accompanied by an increase of the beam energy.

\begin{figure}[h!]
    \centering
    \includegraphics[width=1\linewidth]{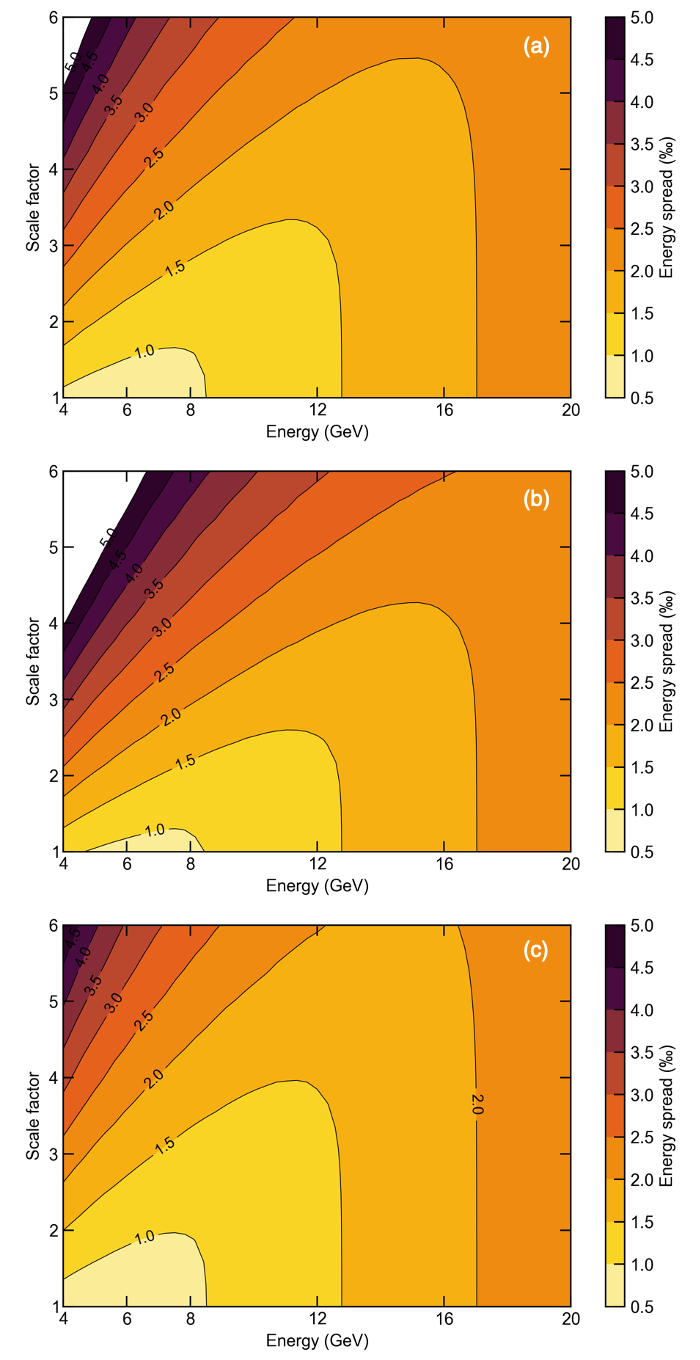}
    \caption{Achievable rms energy spreads as a function of beam energy and 6BA scaling for different coupling ratios: $\kappa = 0.01$ (a), $\kappa = 0.1$ (b), and $\kappa = 0.5$ (c). Equilibrium energy spread is defined by the balance of SR, QE, and IBS. A factor 10 bunch lengthening by a harmonic rf system is assumed. The hatched area denotes the parameters space unavailable due to the space charge tune spread. Bunch charge 1 nC.}
        \label{fig:minimum_energy_spread}
\end{figure}

Finally, we can depict the expected equilibrium energy spread reachable under IBS as it is an important parameter, relevant for the photon science users. Figure~\ref{fig:minimum_energy_spread} presents our numerical findings. Similarly to Fig.~\ref{fig:minimum_emittance} the upper-left part of the parameter space is excluded by SC. In the region not dominated by SC the equilibrium energy spread increases steadily with the energy and does not exhibit a strong dependence on the scaling.

\begin{figure}[h!]
    \centering
    \includegraphics[width=1\linewidth]{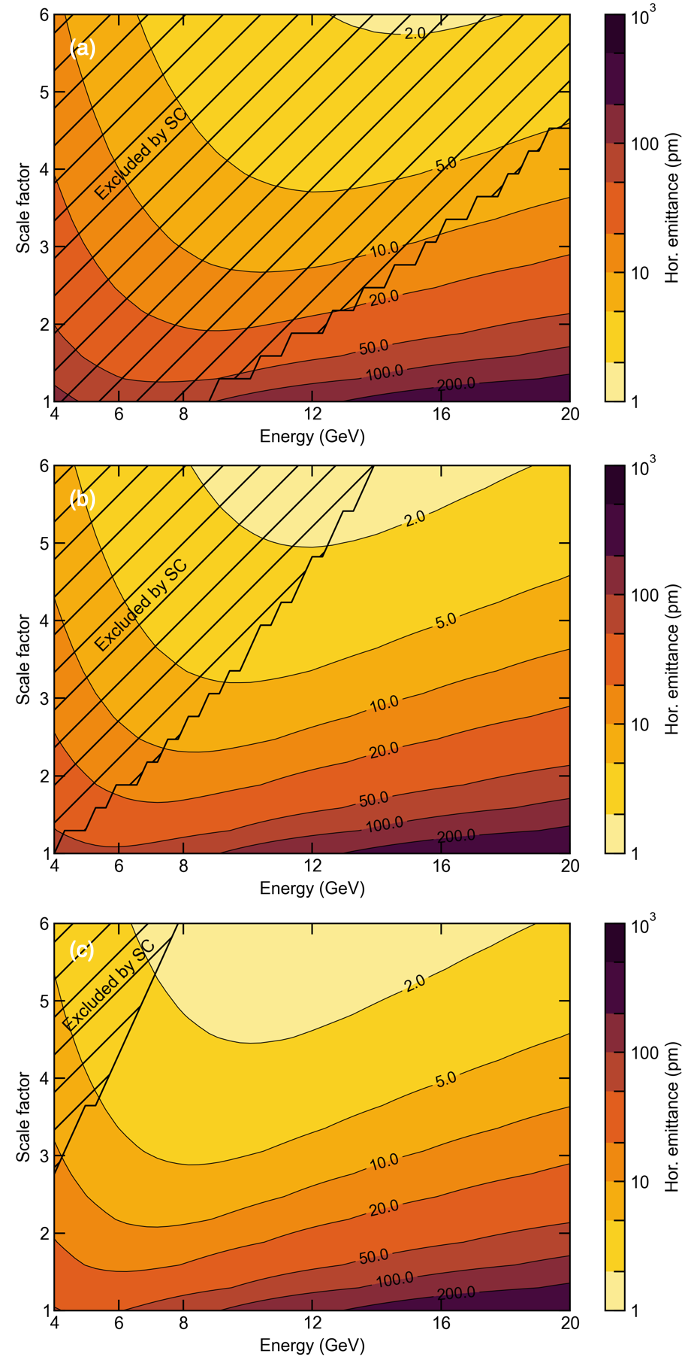}
    \caption{Achievable emittances as a function of beam energy and 6BA scaling for different couplings: $\kappa = 0.01$ (a), $\kappa = 0.1$ (b), and $\kappa = 0.5$ (c). Equilibrium emittance is defined by the balance of SR, QE, and IBS. Bunch stretching by a harmonic rf system to 40~ps for across all cases is assumed. The hatched area denotes the parameters space unavailable due to the space charge tune spread. Bunch charge 1 nC.}
        \label{fig:minimum_emittance_const_length}
\end{figure}

As our assumption on the bunch length might have been overly pessimistic, we also examined another 'extreme' scenario where the bunch length is kept constant between all the cases, at 40~ps. This results only in a minor difference in achievable emittances (Fig.~\ref{fig:minimum_emittance_const_length}). Notably though, the choice of bunch length affects the limits of SC-dominated region with longer bunches allowing reaching smaller emittances at large scaling factors.


Examining the IBS growth rates Eq.~(\ref{eq:ibs_rates}) reveals that they are primarily functions of emittance and energy. At the same time, the SR damping time is independent of emittance and circumference and decreases with the energy. This implies that our scalings should also hold, in principle, for smaller machines, while small emittances might be harder to achieve there in practice. From the point of view of space charge, which depends strongly on the energy and increases with circumference, a higher beam energy and a smaller circumference are preferable.

\subsection{Other effects}

First, let us consider the Touschek lifetime, which depends strongly on beam energy and momentum acceptance:
\begin{equation}
T_{touschek}^{-1} \propto \frac{1}{\gamma^2  \delta_{acc}^3}
\end{equation}
In the state-of-the-art light sources, such as PETRA~IV, this lifetime may reach $\sim 100$~hours. It will further increase with beam energy. At the same time, while emittance reduction might lead to a decrease in momentum acceptance. For a practical lattice, the latter shall remain significantly larger than the rms energy spread (Fig.~\ref{fig:minimum_energy_spread}).


For both single and multi-bunch instabilities the scaling of thresholds is usually inversely proportional to energy. For example, the single-bunch transverse instability limit can be quantified by the TMCI threshold

\begin{equation}
I_{th} \propto \frac{\gamma \omega_s \sigma_z }{\beta_{x,y} Z_T},
\end{equation}
where $Z_{T}$ is the impedance sampled at harmonics of revolution and synchrotron frequency.
In practice this limit is significantly increased with chromaticity, and does not manifest itself unless operating the machine in a mode with few high-charge bunches. At PETRA IV, for the 2 ns uniform bunch filling the single-bunch intensity is roughly a factor 50 smaller than the threshold bunch current. A similar scaling is valid for coupled-bunch instabilities. In general though, coupled-bunch instabilities are not limiting the machine performance when feedback systems are employed. 

 Machine impedance is strongly dependent on choice of apertures, and, in the case of rf, on the resonator frequency and technology. The scaling of geometric impedance is with the inverse second power of vacuum chamber dimension $b^{-2}$, whereas the resistive wall contribution scales with the third power, $b^{-3}$.
 The detailed investigation of impedance goes beyond the present discussion. But both single- and coupled-bunch instability scalings are such that the intensity thresholds generally improve with beam energy and are not limiting the machine performance unless the impedance is drastically increased, which can be a limitation in practice if pursuing high gradient magnet and high-voltage rf through reducing the vacuum chamber dimensions and increasing the rf frequency.



\section{Radiation brightness scaling}
A relevant figure of merit for the undulator radiation is its peak brightness, commonly measured in units of photons/s~mm$^2$mrad$^2$(0.1\% bandwidth
), which can be approximated as
\begin{equation}
    \mathcal{B}_{ph.pk} = \frac{\mathcal{F}_n}{(2\pi)^2\sigma_{T,x}\sigma_{T,y}\sigma_{T,x^\prime}\sigma_{T,y^\prime}}, 
\end{equation}
where $\sigma_T$ include the photon contribution to the effective beam size, $\sigma_{T,x} = \sqrt{\sigma_x^2 + \sigma_r^2}$ and $\sigma_{T,x^\prime} = \sqrt{\sigma_{x^\prime}^2 + \sigma_{r^\prime}^2}$ and similarly for the y-plane, $\sigma_r = \sqrt{\lambda L_u} / 4\pi$. Let us assume operating close to a diffraction limit, such that 
\begin{equation}\label{eq:diff_lim}
    \epsilon_x \sim \epsilon_{D} = \frac{\lambda}{4 \pi}
\end{equation}
Then one may assume that $\sigma_x \sim \sigma_r$ and $\sigma_{x^\prime} \sim \sigma_r^\prime$. 
So, for the sake of an order of magnitude estimate we can write the peak photon brightness as
\begin{equation}\label{eq:peak_brilliance}
    \mathcal{B}_{ph.pk} \sim  \frac{1}{2\pi\sigma_{x}\sigma_{y}} \frac {\partial^2\mathcal{F}_n} {\partial\Theta\partial\psi} \propto \frac {\gamma^2 I} {\epsilon_x^2 \kappa} = \frac {16\pi^2\gamma^2 I} {\lambda^2 \kappa}, 
\end{equation}
where $\kappa$ is a fixed transverse emittance coupling ratio $\kappa = \epsilon_y / \epsilon_x$. Note that the ratio of bunch current to the square of emittance $I / \epsilon_x^2 \propto B$, which is fundamentally limited by space charge tune shift, according to Eq.~(\ref{eq:SC_limit}). This means that the space charge tune shift poses a limit to how far one can push peak photon brightness for a given undulator technology. 
%

Further factors have to be considered. First, the radiation wavelength (first harmonic) of an undulator is
\begin{equation}
 \lambda = \frac{\lambda_u}{2\gamma^2} \left( 1 + \frac{K^2}{2} \right),
\end{equation}
where $K={e B \lambda_u} / {2 \pi m_e c}$. Increasing the beam energy and keeping the same radiation wavelength would require increase of the undulator parameter $K$ or the undulator period $\lambda_u$. Both will result in broadening of radiation spectrum, which is not desirable. Moreover, the brightness improvement with emittance is limited by the diffraction limit on emittance. For a typical radiation wavelength of $10~\text{keV}$, Eq.~(\ref{eq:diff_lim}) yields the diffraction limit of about 10 pm. Decreasing emittance beyond that for this wavelength does not improve brightness. Simultaneous increase in electron beam energy and reduction in emittance would allow to dramatically extend both the diffraction-limited wavelength and the achievable brightness using similar insertion devices.  So, according to~(\ref{eq:peak_brilliance}), a reduction from 20 pm emittance to 2 pm and increasing energy to 18 GeV from 6 GeV leads to an order of magnitude in photon energy reach and diffraction limit, and in nearly three orders of magnitude in peak brightness. 

\begin{figure}
    \centering
    \includegraphics[width=1\linewidth]{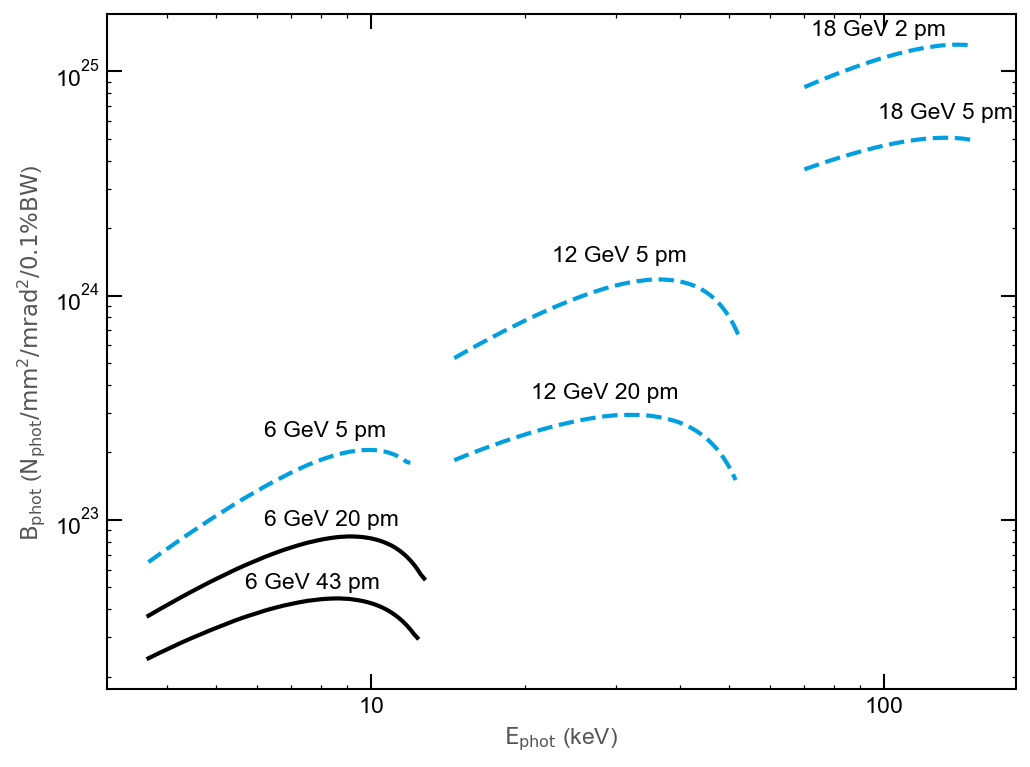}
    \caption{Radiation brightness for various scenarios. For all curves insertion device length is $L=4.5$ m,  $\beta_x=\beta_y= 2$~m, $\kappa=0.1$. For 6 GeV: $\lambda_u$ = 2.3 cm, $\sigma_p=0.7\times10^{-3}$ for 43 pm emittance and  $\sigma_p=1\times10^{-3}$ for 20 pm emittance. For 12 GeV: $\lambda_u$ = 2.3 cm, $\sigma_p=1.5\times10^{-3}$. For 18 GeV:$\lambda_u$ = 1.3 cm, $\sigma_p=2\times10^{-3}$. }
        \label{fig:brilliance}
\end{figure}

Brightness for given undulator and beam parameters can be found more precisely with the help of specialized codes. Calculations of brightness with SPECTRA \cite{spectra} for several scenarios of emittance reduction and energy increase are shown in Fig.~\ref{fig:brilliance}. There the solid black curves represents the parameters achievable at PETRA IV with and without taking damping wigglers into account. Even a reduction to 5 pm emittance, which already beyond the space charge and IBS limitation~(Fig.~\ref{fig:minimum_emittance}), would increase the brightness by no more than a factor 3 for a typical insertion device with the parameters listed in the caption of Fig.~\ref{fig:brilliance}. Other dashed curves represent hypothetical scenarios for simultaneous increase of beam energy and decrease of emittance based on the H6BA scaling. At about 18 GeV the theoretically achievable emittance of 2 pm (c.f. Fig.~\ref{fig:minimum_emittance}) would result in simultaneous increase of brightness by more than two orders of magnitude and of photon energy reach by one order of magnitude.

\section{Design implications for high-energy low-emittance sources}

The dramatic increase of brightness while pursuing lower-emittance lattices for higher beam energy comes at a cost. Consider a scaling of the number of bends in the PETRA~IV lattice by a moderate factor 2 and increase of energy from 6 to 8 GeV which would result in achievable emittance of about 10 pm (see Figure \ref{fig:minimum_emittance}). Assuming usage of permanent magnet technology and reducing spacing between magnets such that the sextupoles and quadrupoles could be made twice longer compared to the resistive magnet design, we would arrive at quadrupole gradients of about 300-400 T/m, which is technically possible with permanent magnets~\cite{Vannozzi_2020}, and sextupole strength of about 80000 $\mathrm{T/m^2}$, which is beyond the current technological capabilities. A novel lattice concept beyond hybrid multi-bend achromat is required for chromaticity compensation in this case, with possibly more distributed sextupole compensation. Both vacuum pumping and impedance issues will become more prominent problems in such designs.

Whatever the approach for lowering the emittance further, dynamic aperture limitations will become more dramatic, with values possibly below 1 mm. A combination of novel injection schemes with future low-emittance injectors such as the LPA injector being developed at DESY \cite{Winkler:2025kzz, Antipov:2021eko}, would allow to operate such machines efficiently.

HOM-damped normal-conducting cavities used in light sources would be inefficient to cope with an increased voltage required to compensate synchrotron radiation losses at a higher beam energy. Here, superconducting CW cavities can be employed more efficiently \cite{Posen:2021zzz, Chen:2025iry}.

\section{Conclusion and outlook}

We analyzed potential limitations in further emittance reduction in MBA lattices on the example of PETRA and showed that space charge effects and IBS present a fundamental limitation on the way to achieving emittances below about 10 pm unless the beam energy is raised. This limitation of emittance naturally translates to the photon brightness, where a significant improvement without raising the beam energy is impossible. On the other hand, increasing the energy from 6 to 16-18 GeV would open a completely new field of opportunities, increasing the brightness by more than two and the photon spectral reach by one order of magnitude. Technological limitations stand in the way of achieving this improvement with the known multi-bend achromat lattice concepts, and further R\&D into design of low emittance lattices is required.

The limitations are independent of the machine size. However, smaller emittances are more accessible at larger machines. This make photon science applications an appealing complementary option to the high energy physics program at the Future Circular Collider complex~\cite{Agapov:2928809}. There an emittance of about 1-2 pm could be reached in the booster of 90 km circumference at 20 GeV with standard technology before encountering intensity limitations.

\section{Acknowledgments}
We thank Gianluca Geloni for cross-checking the brightness calculations, Frank Zimmermann for discussions concerning FCC booster parameters, and Reinhard Brinkmann for carefully reading the manuscript and making useful suggestions for improvement.

\bibliography{main}

\appendix
\section{Appendix A: Space charge tune shift approximation for light sources}\label{ap:A}

Consider a typical scenario of a machine operating with flat beams, $\sigma_y \ll \sigma_x$, then the largest tune shift occurs in the vertical, $y$-plane. 
Further, assume that the ring is designed with a low-emittance approach in mind and the dispersion contributions to the beam size are, on average, small. Then the beam size is mostly governed by its emittance $\epsilon$ and one can substitute in Eq.~(\ref{eq:SC_general_case}) $\sigma_{x,y} = \sqrt{\beta_{x,y}\epsilon_{x,y}}$, arriving at
\begin{equation}
    \Delta \nu^{SC}_{y} = - \frac{N r_e C}{(2\pi)^{3/2}\gamma^3\sigma_z} \biggl \langle
    \frac{\beta_{y}}{\sqrt{\beta_y\epsilon_y}(\sqrt{\beta_x\epsilon_x}+\sqrt{\beta_y\epsilon_y})} 
    \biggr \rangle.
\end{equation}
Introducing the coupling ratio as $\kappa = \epsilon_y / \epsilon_x$ the expression can be rewritten as
\begin{equation}
    \Delta \nu^{SC}_{y} = - \frac{N r_e C}{(2\pi)^{3/2}\gamma^3\sigma_z\epsilon_x\kappa}
    \biggl \langle 
    \frac{\sqrt{\kappa} \sqrt{\beta_{y}/\beta_x}}{(1+\sqrt{\kappa}\sqrt{\beta_y / \beta_x})} 
    \biggr \rangle,
\end{equation}
and for small $\kappa$ the term in the brackets is approximately
\begin{equation}
    \biggl \langle 
    \frac{\sqrt{\kappa} \sqrt{\beta_{y}/\beta_x}}{(1+\sqrt{\kappa}\sqrt{\beta_y / \beta_x})} 
    \biggr \rangle
    \approx \biggl \langle
    \sqrt{\kappa} \sqrt{\frac{\beta_{y}}{\beta_x}}\left(1-\sqrt{\kappa}\sqrt{\frac{\beta_y}{\beta_x}}\right)
    \biggr \rangle.
\end{equation}
The last product of a type $p\times (1-p)$ is well-known. As long as $0 < p < 1$ it cannot exceed 1/4. Therefore, we can replace the average with its upper limit, 1/4. With this simplification the final estimate of the vertical tune shift becomes
\begin{equation}
    \Delta \nu^{SC}_{y} \approx - \frac{1}{4\kappa}\frac{N r_e C}{ (2\pi)^{3/2}\gamma^3\sigma_z\epsilon_x}.
\end{equation}

The formula above only applies to small coupling ratios and has to be modified slightly for the case of $\kappa \sim 1$ \cite{IPAC-2025-SC}, but the general scaling remains the same.

In order for the particle motion to be stable this tune shift shall never cross an integer resonance. Considering that the light sources typically operate with their fractional tunes below 0.5 one can assert $|\Delta \nu^{SC}_{y}| \lesssim 1/2$.  Then introducing the peak beam brightness as $B = 2 I / \pi^2 \epsilon_x \epsilon_y =  2 I / \pi^2 \epsilon_x^2 \kappa$, where $I$ stands for the peak bunch current, one then arrives at Eq.~(\ref{eq:brighness_limit}).

\end{document}